\documentclass[aps,prl,footinbib,tightenlines,twocolumn,floatfix, superscriptaddress]{revtex4-1}

\usepackage{graphicx}
\usepackage{amsmath}
\usepackage{amssymb}
\usepackage{hyperref}
\usepackage{soul}
\usepackage[pdftex]{color}
\usepackage{subfigure}

\newcommand{\sgn}[1]{\mathrm{sgn} \left( #1 \right)}

\newcommand{\be}{\begin{eqnarray}}
\newcommand{\ee}{\end{eqnarray}}
\newcommand{\bbm}{\begin{bmatrix}}
\newcommand{\ebm}{\end{bmatrix}}
\newcommand{\bpm}{\begin{pmatrix}}
\newcommand{\epm}{\end{pmatrix}}

\begin{document}

\title{Bulk-boundary quantum oscillations in inhomogeneous Weyl semimetals}
\author{Dmitry I. Pikulin}
\affiliation{Station Q, Microsoft Research, Santa Barbara, California 93106-6105, USA}
\author{Roni Ilan}
\affiliation{Raymond and Beverly Sackler School of Physics and Astronomy, Tel-Aviv University, Tel-Aviv 69978, Israel}

\begin{abstract}
Weyl fermions in an external magnetic field exhibit the chiral anomaly, a non-conservation of chiral fermions. In a Weyl semimetal, a spatially inhomogeneous Weyl node separation causes similar effect by creating an intrinsic pseudo-magnetic field with an opposite sign for nodes of opposite chirality. In the present work we study the interplay of external and intrinsic fields. In particular, we focus on quantum oscillations due to bulk-boundary trajectories. When caused by an external field, such oscillations are a proven experimental technique to detect Weyl semimetals. We show that the intrinsic field leaves hallmarks on such oscillations by decreasing the period of the oscillations in an analytically traceable manner. The oscillations can thus be used to test the effect of an intrinsic field and to extract its strength.

\end{abstract}
\maketitle

Band structure of a three-dimensional crystalline material can exhibit non-degenerate band crossings (nodes) in momentum space around which the Berry curvature behaves as a magnetic field emerging from a monopole. These sources or sinks of the ``Berry flux'' are known as the Weyl nodes, and must come in pairs of opposite sign in the Brillouin zone. Fermions in the states around such nodes can be assigned a quantum number, chirality, determined by the charge of the ``Berry monopole''. The total chirality in the Brillouin zone must be zero. The fermions around the nodes are relativistic, and their low energy dynamics is described by the Weyl equation. The materials are thus termed Weyl semimetals~\cite{volovik_book,Chiu2016review,burkov2011topological,wan2011topological,turner2013beyond,armitage2017weyl}. As such, they can realize an old concept from quantum field theory - chiral anomaly\cite{Peskin}.

The chiral anomaly is the non-conservation of chiral charge in relativistic quantum field theory. The consequences of the existence of the chiral charge in the field theory are well known and celebrated. In particular, the chiral magnetic effect (CME) is a current response generated along the direction an externally applied magnetic field~\cite{Peskin, Fukushima2008}. The same phenomenon is expected to occur in WSM. The observation is complicated by the Neilsen-Ninomiya theorem, which constraints the chiral charges to come in pairs of opposite chirality on the lattice. Therefore the CME cannot exist in a crystal in equilibrium due to a cancellation of contributions from nodes of opposite chiralities~\cite{NielNino81b,Nielsen1983,vazifeh2013electromagnetic,ma2015chiral}. Nevertheless, one can obtain a CME in non-equilibrium conditions, either through dynamics, or through an imbalance of chiral chemical potentials. The latter is possible to generate by applying parallel electric and magnetic fields. The resulting conductivity enhancement has been observed in experiments measuring magnetoresistance\cite{Son:2013jz,Kim2013,burkov2015negative,huang2015observation,xiong2015evidence,li2016chiral,arnold2016negative}. 

Recent works have discussed novel ways for creating a CME in WSMs that can exist in equilibrium. The trigger for such a CME is an intrinsic pseudo-magnetic field and can be sustained locally while vanishing when integrating over the entire sample~\cite{pikulin2016chiral,grushin2016inhomogeneous,o2017superconductivity}. This phenomenon is linked to a known fact from classical electromagnetism -- bound currents as a result of inhomogeneous magnetization can flow within a medium so long as the total dissipation-less current vanishes when averaged over the volume of the sample. If the Weyl node separation is stemming from the magnetization, then the magnetization drives the CME. 

In this work we address an outstanding challenge concerning the physics of
WSM by proposing a direct way to detect and quantify the emergence of pseudo-magnetic fields and the resulting pseudo-CME. We achieve this by exploring the effects of inhomogeneities on a known and proven experimental scheme: quantum oscillations due to semiclassical trajectories traversing the bulk and surface of WSM~\cite{Potter:2014vz,zhang2016quantum,moll2016transport}. This is a striking transport measurement carried out on WSM and exemplifying the existence of bulk-surface trajectories that result in a coherent periodic motion driven solely by the external magnetic field. In the present work we show that pseudo-magnetic fields can bend the bulk quasi-particle trajectories and hence have immediate and quantifiable effects on the period of such oscillations. 

The emergence of pseudo-fields in Dirac materials has striking manifestations in graphene and WSMs. In WSMs it has been recently been claimed to play a key role both in the understanding of the physics of Fermi arcs, as well as in unraveling an equilibrium CME.  In graphene, lattice deformations couple to the electronic degrees of freedom as gauge potentials that do not break time-reversal symmetry, but nevertheless result in the formation of Landau levels~\cite{Vozmediano2010109,LBM10}. Preserving time-reversal symmetry comes about through the coupling of the pseudo-fields with an opposite sign to the two valleys of graphene. The situation in WSM is similar, albeit richer. Pseudo-fields can emerge via lattice deformation as well as inhomogeneous magnetization, and the emergent pseudo-gauge fields couple to fermions of opposite chirality with an opposite sign~\cite{liu2013chiral,chernodub2014condensed,shapourian2015viscoelastic,cortijo2015elastic,sumiyoshi2016torsional,
pikulin2016chiral,grushin2016inhomogeneous,chernodub2017chiral,nica2017landau}. The resulting Landau level structure from spatial inhomogeneities is similar in many ways to the one resulting from external fields: it generates a lowest Landau level that disperses along the direction of the field near each node. But it is crucially different in one respect: the chirality of the lowest Landau level is indifferent to the sign of the node. The origin of the pseudo-CME in Weyl semimetals is rooted in the fact that while the usual CME is zero in equilibrium due to the cancellation of currents from nodes of opposite chirality, there is no such local cancellation for pseudo-fields. Therefore, for a time-reversal broken WSM with two nodes there will be a net local CME. Inversion-broken systems with a minimum of four nodes have counter-propagating currents stemming from time reversed pair of nodes. 

 The principle behind the effect of pseudo-fields on quantum oscillations is simple. Quantum oscillations stem from trajectories that traverse the bulk via the dispersion of the lowest Landau level, combined with a semiclassical sliding motion along the arcs at the surface perpendicular to the direction of the field~\cite{Potter:2014vz,zhang2016quantum}. When fixing the direction of the external magnetic field such trajectories are deformed due to intrinsic magnetic fields. Particles in the bulk are forced to move in the direction of the total effective magnetic field felt by the Weyl node, which is a superposition of the two components (external and intrinsic). 
 
 Below we derive the relevant formula for the period of the semiclassical oscillations. As we show, deformed trajectories have a strong quantifiable effect on the density of states as well as the frequency of oscillations in experimentally available responses (e.g. conductivity). The striking difference as compared to oscillations emerging purely from the external fields is that while oscillations in the absence of pseudo-fields are periodic and depend only on the total momentum space enclosed by the Fermi arcs, with bulk pseudo-fields the interval between oscillations becomes field-, pseudo-field-, as well as thickness-dependent. We support our predictions with numerical simulations performed using a tight-binding model compatible with the physics of $\text{Cd}_3\text{As}_2$, a Dirac semimetal on which the original quantum oscillations experiment was performed~\cite{moll2016transport}.

\begin{figure}[t]
\begin{center}
\includegraphics[width=.3\textwidth, angle =-90]{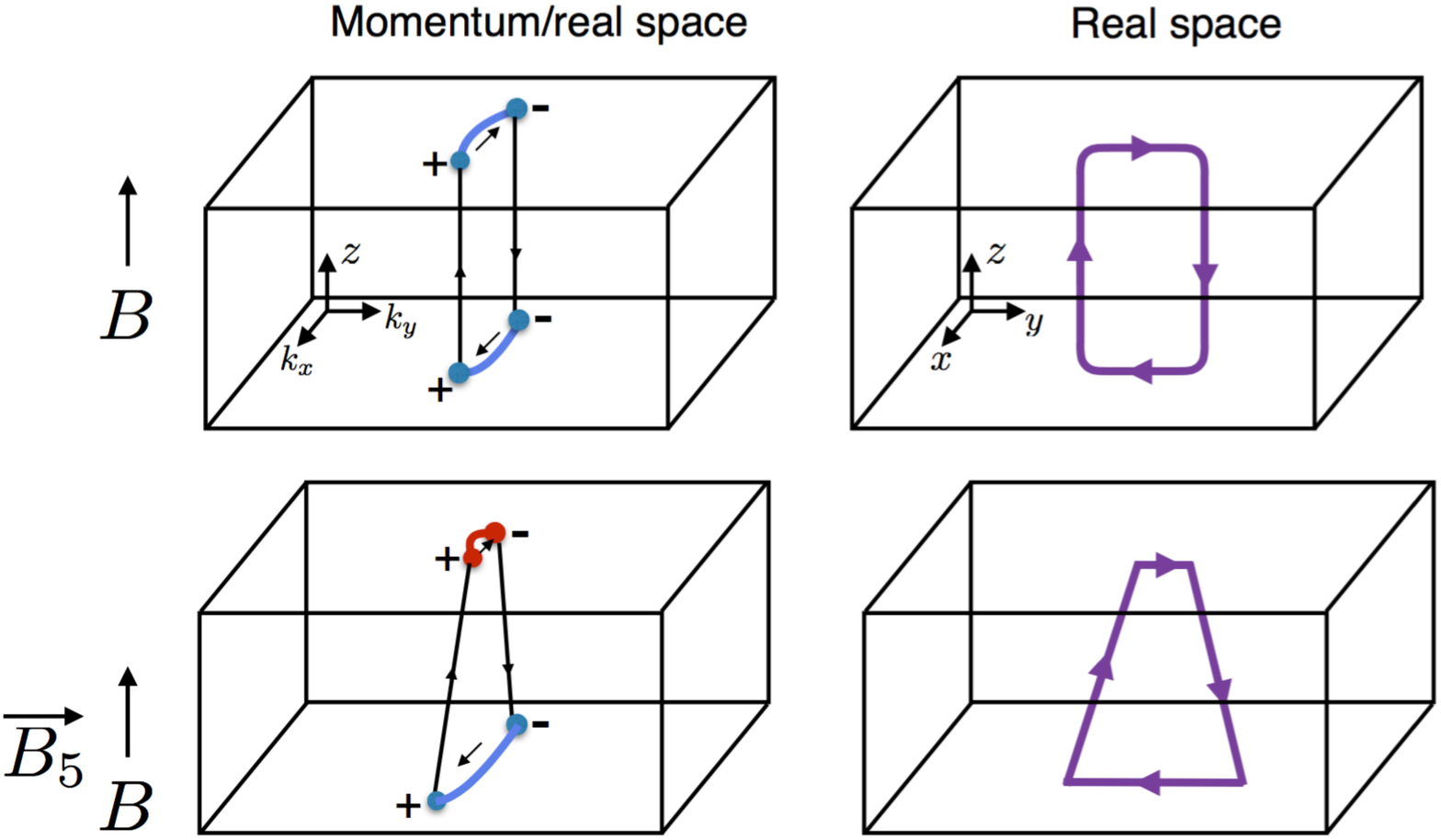}
\end{center}
\caption{Semiclassical closed trajectories that produce quantum oscillations. The upper left panel describes the mixed momentum space and real space picture, where the motion in the bulk is shown in real space and follows the direction of the externally applied magnetic field. The surface trajectories are illustrated in momentum space of the surface plane, where particles' trajectories drift along the Fermi arc to exchange chirality before sinking back into the bulk. The upper right panel presents the real space map of the trajectory. In the lower left panel the mixed trajectories are shown again in the presence of a pseudo-field $B_5$, perpendicular both to the nodal separation as well as the external field. The top and bottom arcs have different lengths and the bulk trajectories are tilted. On the right the deformed real space trajectory is shown.} \label{WSM_E}
\end{figure}

To make our discussion concrete, consider a film of a WSM. For simplicity, we consider a WSM with a single pair of Weyl nodes, but the analysis straightforwardly generalizes. We take the Weyl node separation $\textbf{p}_0$ to be along $p_y$ as depicted in Fig~\ref{WSM_E}. Then the low energy Hamiltonian is 
\begin{equation}\label{WeylH}
H_0=\pm v (\textbf{p}\pm\textbf{p}_0)\cdot\sigma - \mu_0
\end{equation}
Where $v$ is the velocity, which we take here to be isotropic (the case of anisotropic velocity is discussed in the supplementary material), and $\mu_0$ is the chemical potential offset with respect to the Dirac point. We note that written in this form, $\textbf{p}_0$ couples to the Hamiltonian as an axial vector potential. When $\textbf{p}_0$ is a constant, its importance is in the separation of nodes but beyond that it brings about no other interesting structure since $\nabla\times\textbf{p}_0=0$. This is in accordance with the vector potential interpretation. Strain renders the Weyl node separation space-dependent~\cite{cortijo2015elastic,RKY16,pikulin2016chiral,grushin2016inhomogeneous}, and may make $\nabla\times\textbf{p}_0$ non-zero. For simplicity we consider a strain profile that makes $\textbf{p}_0$ depend linearly on the $z$ coordinate. As we show below, such profile corresponds to a physical strain configuration. In such case we can write the Weyl nodes separation as $\textbf{p}_0(z)=(b_0-B_5z)\hat{y}$
Now, taking the curl of $\textbf{p}_0(z)$ we can define $\textbf{B}_5=1/e\nabla\times \textbf{p} _0(x)=B_5\hat{x}$, which is a pseudo-magnetic field that couples to Weyl nodes of opposite chirality with an opposite sign. Therefore,  this position-dependent Weyl node separation and, as a result, bulk strain leads to intrinsic pseudo-magnetic fields. The magnetic field breaks the linear Dirac spectrum around each node into Landau levels, with a dispersion relation 
\begin{eqnarray}\label{LL}
&&\epsilon_n(k_x)=\pm\sgn{n} v\sqrt{k_x^2+2\ell^{-2}_{B_5}}\\
&&\epsilon_0(k_x)=\sgn{B_5}v k x
\end{eqnarray}
Here and later $\ell_{B(B_5)} = \frac{1}{\sqrt{e B (B_5)}}$, $\hbar=1$. As the pseudo-magnetic field acts oppositely in the two Weyl nodes, for the given node configuration the chirality of the lowest Landau level is the product of the sign of $B_5$ and corresponding Weyl node chirality. 

We now turn to the influence of $B_5$ on the bulk trajectories. Assuming an externally applied field in the $z$ direction, $\textbf{B}=B\hat{z}$ the total field experienced by a particle with chirality $s$ (with $s=\pm1$) is 
\begin{eqnarray}
&&\textbf{B}_s=\textbf{B}+s\textbf{B}_5=B\hat{z}+s B_5\hat{x}.
\end{eqnarray} 
The intrinsic field thus tilts the bulk trajectories from the direction of the external field by an angle 
$
\theta=\tan^{-1}B_5/B
$.
Hence, the bulk path traversed is of length 
$
L'=L/\cos{\theta}=L\sqrt{1+(B_5/B)^2}.
$

 In order to determine the period of quantum oscillations, we follow the analysis in Ref.~\cite{zhang2016quantum}, and derive the phase space quantization condition for the closed quasiparticle trajectories 
\begin{equation}
\oint_c\textbf{p}\cdot \textbf{dr}=2\pi (n+\gamma)
\end{equation}
where $\gamma$ is a constant offset. 
According to the discussion above, the integral for the mixed bulk-surface trajectories is broken into two pieces, due to the presence of the intrinsic and external vector potentials, namely the integral is taken over four segments of the trajectories, including the two arcs, and two bulk branches linking the top and the bottom surfaces. See Fig. \ref{WSM_E} for depiction of the trajectories in the mixed real-momentum and purely real spaces. For the arcs, the integral yields
\begin{equation}
\int\textbf{p}\cdot \textbf{dr}=e\Phi
\end{equation}
Where $\Phi$ is the total flux enclosed by the real space orbit in the surface plane. If the surface encircled is $S$, then 
\begin{equation}
\Phi=SB=BS_k\ell_B^4
\end{equation}
with $S_k$ the momentum space area enclosed by the arcs. At small chemical potential this area is approximately given by $S_k=k_0(\mu+\mu_0)/v$, where $k_0$ is the total length of the arcs, $\mu$ is the chemical potential measured from the Weyl nodes, $\mu_0$ is the chemical potential offset as discussed in Ref.\cite{zhang2016quantum} and $v$ the Fermi velocity at the surface which we take to be equal to that of the bulk. Note that in principle, $k_0$ may depend on $B_5$: the presence of $B_5$ in the bulk necessarily means the length of the two arcs on opposite surfaces is inequivalent. Here, we will analyze the simplest case in which strain enhances the arc length on one surface by the same amount it shortens the arc on the opposite surface. Then, the total length of the surface trajectory is not modified by $B_5$, although $B_5$ is finite in the bulk. This corresponds to the physical strain, corresponding to bending the Weyl semimetal field, discussed in \cite{Liu2017}. In the supplementary material we discuss other cases where changes in the two arcs do not compensate one another.

\begin{figure}[t]
\includegraphics[width=\columnwidth]{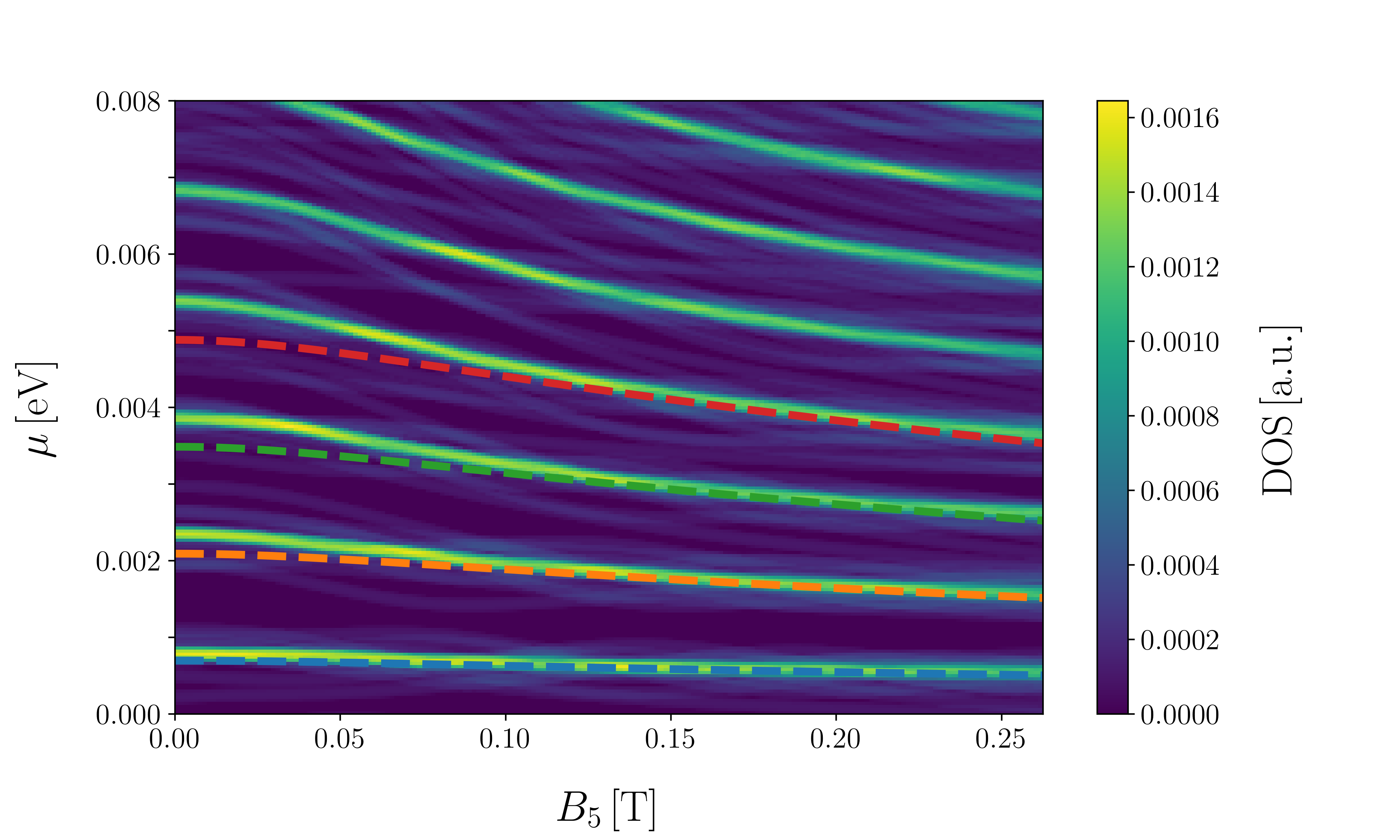}
\caption{Density of states of the single spin block of particle-hole symmetric Cd$_3$As$_2$ under both external magnetic field and stress-produced pseudo-magnetic field. Parameters of the simulations are: $e_s = e_p = 0.0574$eV, $m_{s\perp} = m_{p\perp} = 9.014$eV nm$^2$, $m_{s\parallel} = m_{p \parallel} = 6.407$eV nm$^2$, $A = 1.21287$eV nm. The simulation is performed on a cubic lattice with lattice constant $a=8$nm, the thickness of the material is $240$nm, and the width of the stripe is $480$nm. External magnetic field is $B=5$T. Horizontal scale shows the effective pseudomagnetic field, and vertical is the energy as measured from the Weyl nodes. Striped lines are expressions from \eqref{eq:energies} \textit{without a free parameter}\label{fig:B5}}
\end{figure}

\begin{figure*}[t!]
\centering
\subfigure[]{
\includegraphics[width=5.5cm]{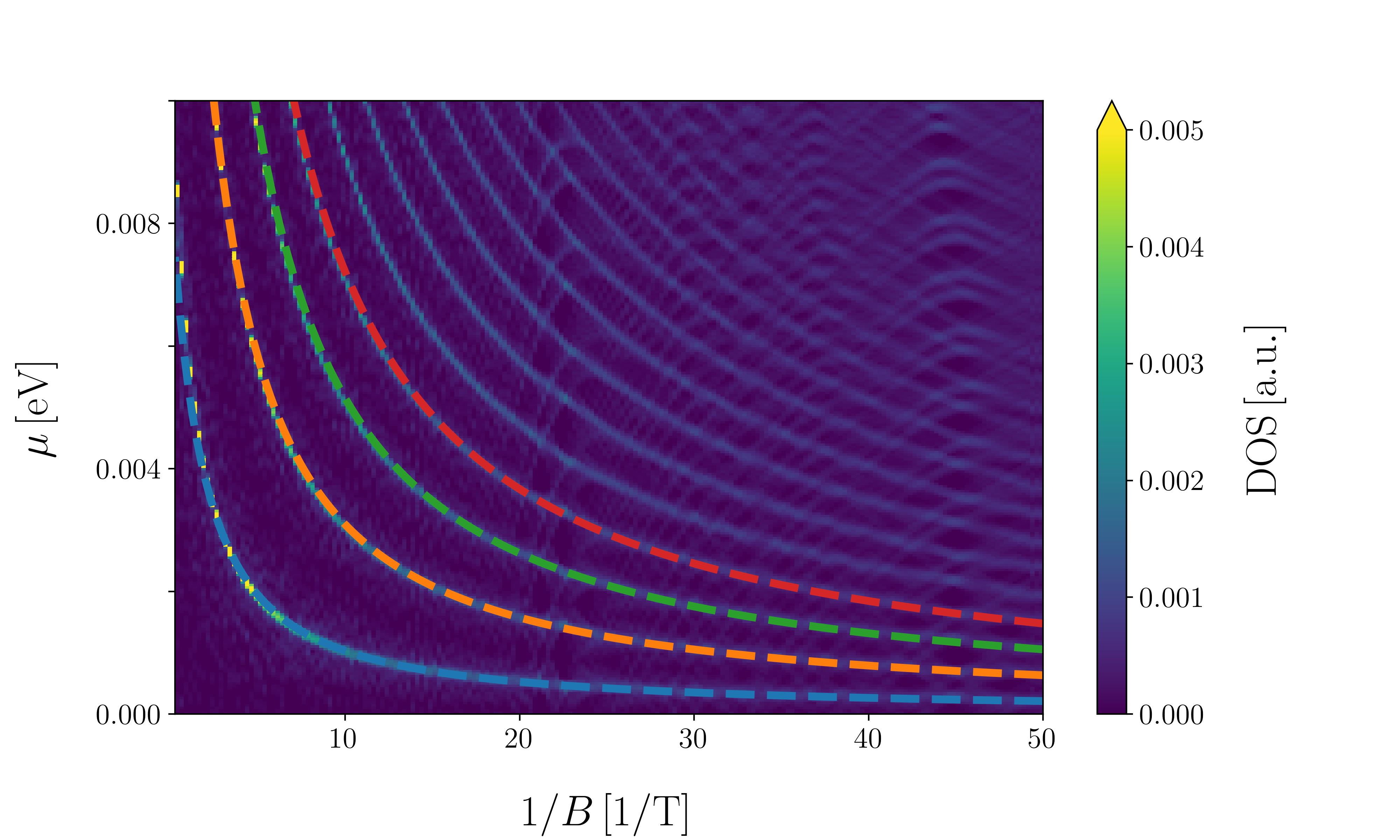}}
\subfigure[]{
\includegraphics[width=5.5cm]{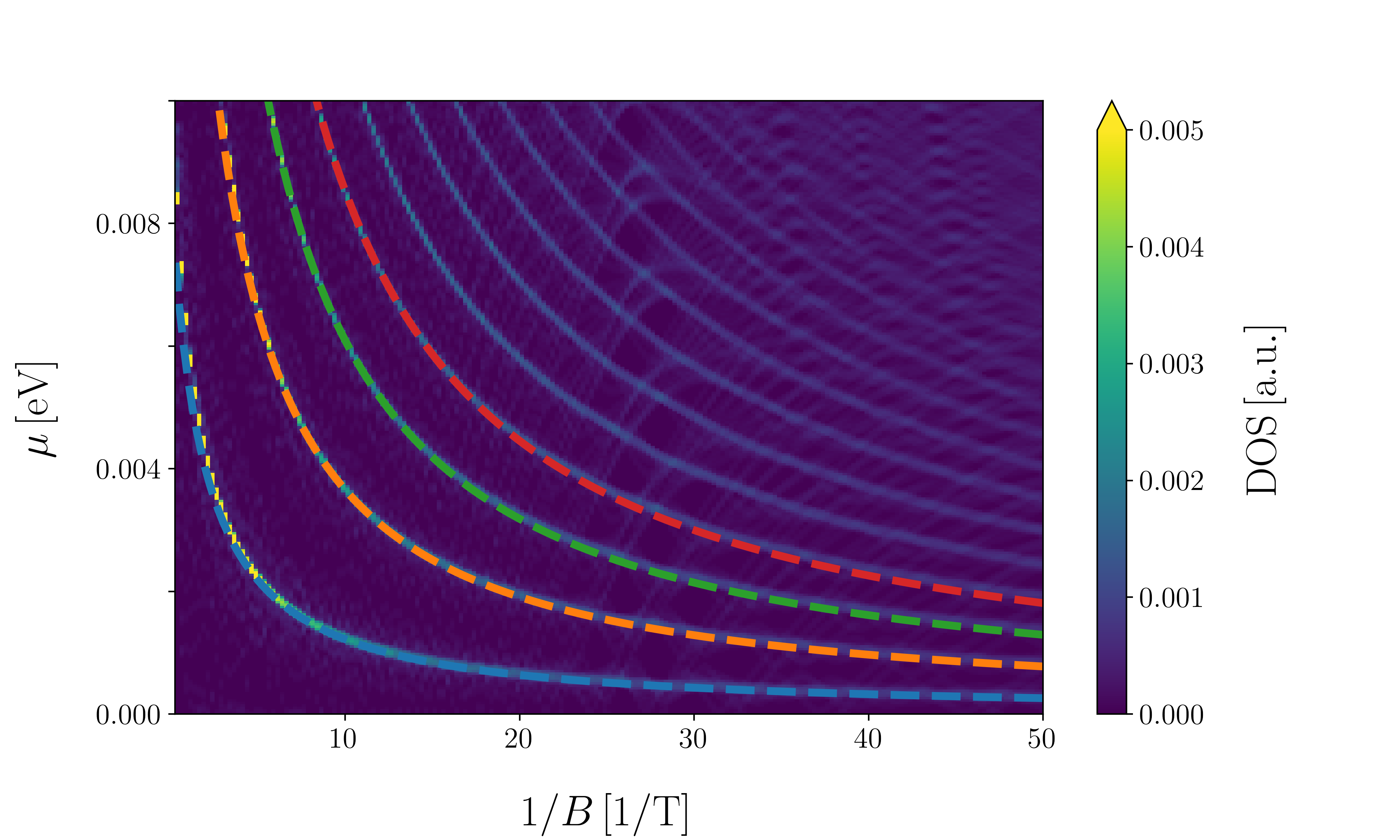}}
\subfigure[]{
\includegraphics[width=5.5cm]{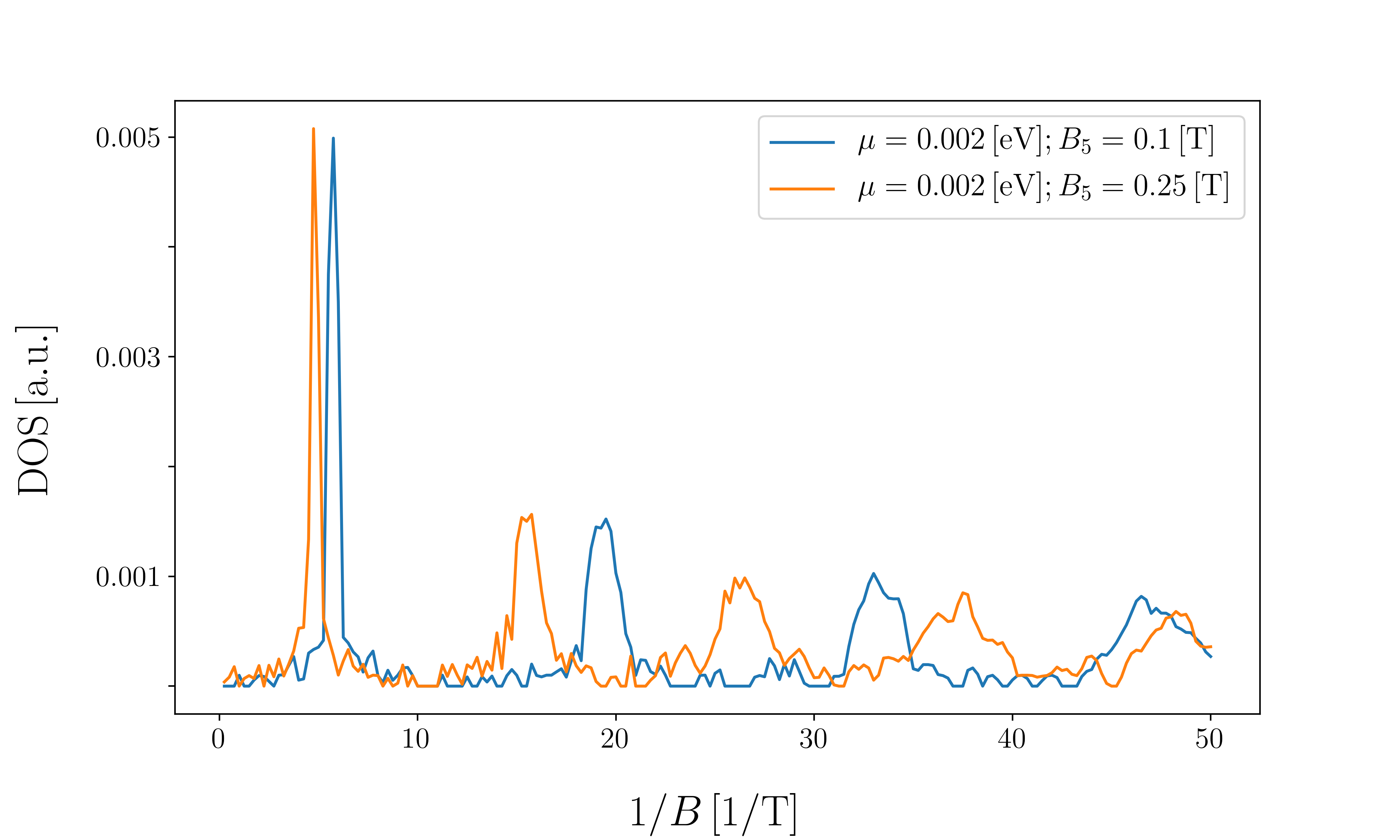}}
\caption{(a-b): Same as Fig. \ref{fig:B5}, but now variying $1/B$ on the horizontal axis, while keeping $B_5$ constant. (a): $B_5=0.25$T, and (b): $B_5=0.1$T. (c): linecuts of (a) and (b) at $\mu=0.002$eV, showing that the periodicity of the quantum oscillations gets shifted with $B_5$. \label{fig:B}}
\end{figure*}

In the bulk, the trajectory of the particles is parallel to the total magnetic field, so that~\footnote{Generically, the integral $\int\textbf{p}\cdot \textbf{dr}$ will also contain another contribution from the dot product of the nodal separation vector $\bf{p}_0$ and the external field (see Ref.~\cite{zhang2016quantum}), however, for our choice of directions, this contribution is zero.  In addition, the contribution proportional to $\int \bf{A}\cdot\textbf{dr}$  also vanishes, as the bulk trajectories enclose no net flux.} 
\begin{equation}
\int\textbf{p}\cdot \textbf{dr}=L\sqrt{1+(B_5/B)^2}(2\mu/v).\label{bulk}
\end{equation}
Defining $L_{eff}=2L\sqrt{1+(B_5/B)^2}$ and summing the two contributions together we have 
\begin{equation}\label{semiclassic}
2\pi(n+\gamma)=\mu L_{eff}/v+eBS_k\ell_B^4.
\end{equation}

From equation \eqref{semiclassic} we can obtain our first testable prediction. The positions of the bulk-boundary energy levels, given by:
\begin{equation}
\epsilon_n=\frac{2\pi(n+\gamma)v - k_0 \mu_0 l_B^2}{L_{eff}+k_0\ell_B^2} \label{eq:energies}
\end{equation}
and are strongly affected by $B_5$. Increasing $B_5$ makes the levels more dense. Furthermore, we can consider quantum oscillations as a function of $B$ or $B_5$. At $B_5=0$, the oscillations' period is $\Delta(1/B)\approx 2\pi e/S_k$. As a small $B_5$ is introduced, a correction is added to the denominator, $S_k\rightarrow S_k+2\mu eLB^2_5/vB$, making the oscillations non-periodic, as the separation between peaks becomes magnetic field dependent. Moreover, as opposed to the case of a purely external magnetic field, the separation between peaks is now thickness dependent. In the opposite limit $B_5\gg B$, we obtain $\Delta(1/B)= 2\pi e/(S_k+2\mu L B_5/v)$, from which it is clear that that while oscillations are periodic in $1/B$, $B_5$ decreases the period of oscillations, and makes it depend on the sample thickness.

Thus we obtain our main experimental predictions: closed bulk-boundary trajectories produce peaks in DOS at energies corresponding to the solutions of the Eq. \eqref{eq:energies}. These can be observed in conductance (Shubnikov-de Haas, SdH) and magnetization (de Haas-van Alphen, dHvA).

\textit{Numerical tests.} To confirm the validity of the results above and their applicability to realistic materials and conditions we performed numerical simulations of a Hamiltonian applicable to Cd$_3$As$_2$ and Na$_3$Bi Dirac semimetals. In these semimetals we can neglect the spin-orbit coupling, thus we use the basis of a single spin, $|s\uparrow,p\uparrow\rangle$:
\begin{align}
H(k) = \begin{pmatrix}
E_s & A p_+\\
A p_- & E_p 
\end{pmatrix},
\end{align}
where:
\begin{align}
E_s &= e_s + m_{s\perp} p_x^2 + m_{s\parallel} p_\parallel^2; \\ E_p &= e_p + m_{p\perp} p_x^2 + m_{p\parallel} p_\parallel^2,
\end{align}
$p_\pm = p_y \pm i p_z$, $p_\parallel = (p_y, p_z)$, and the parameters used are summarized in Fig. \ref{fig:B5}. Note that we use the particle-hole symmetric version of the model for simplicity ($E_p = - E_s$). For this model we identify: distance between the Weyl nodes $\mathbf{p}_0=(\sqrt{\frac{e_s}{m_{s\perp}}}, 0, 0)$, and velocities around the Weyl points, $v_\perp = 2 \sqrt{e_s m_{s\perp}}$, and $v_\parallel = A$. For the purpose of our simulations we set $v_\perp=v_\parallel$ by changing $A$. This makes comparison to \eqref{eq:energies} straightforward.
 
We use the same procedure as in \cite{Liu2017} to introduce the $B_5$ field according to the displacement vector:
\begin{align}
\mathbf{u} = (2\alpha x z, 0, 0),
\end{align}
where $\alpha$ controls the strength of the strain. From this we compute the elements of the symmetric strain tensor $u_{ij} = (\partial_i u_j + \partial_j u_i) / 2$. Then $u_{13} = 2 \alpha z$, and $u_{11} = 2 \alpha x$ and correspondingly the pseudomagnetic field generated by the strain. In this model the $u_{31}$ has much smaller contribution to the pseudomagnetic field than $u_{11}$ due to a small prefactor $1/(a p_0)^2$, where $a$ is the lattice constant of the material. In Cd$_3$As$_2$ this prefactor is $\approx 1/57$ \cite{Liu2017}. We thus only use $u_{11}$, which gives uniform pseudo-magnetic field in $y$ direction of strength $B_5 = 2\alpha \frac{\hbar c}{e a} \cot a p_0$. Such strain corresponds to the hopping modification according to:
\begin{align}
t_x \tau_z \to t_x (1 - 2 \alpha z) \tau_z.
\end{align}
Such modification makes the distance between the Weyl nodes, set by hopping in $z$ direction, position-dependent, in accordance with the definition of $B_5$ we used above. 

To introduce a real magnetic field we use the standard Peierls substitution 
\begin{align}
t_x \to e^{i B y a / (h / e)} t_x,
\end{align}
which produces a real magnetic field in the $z$ direction. With both real and pseudo-magnetic field present only $x$ direction remains infinite in the simulations. Thus, even though the obtained agreement with the theory seen in Fig. \ref{fig:B5} and Fig. \ref{fig:B} is very good, we could not get rid of the finite-size effects completely. 

In our numerical results we show DOS of a slab of the Cd$_3$As$_2$ for a fixed $B$($B_5$), while varying $B_5$($B$) correspondingly. This allows us to model the two experimental scenarios. We imagine putting a sample into fixed external field and continuously bending it to create the pseudo-magnetic field (see Fig. \ref{fig:B} for the change in DOS, corresponding to this scenario). Alternatively, one can fix the bend of the sample and change the external field (see Fig. \ref{fig:B5} for similar results in this case). We show the result of the equation  \eqref{eq:energies} \textit{without fitting parameters} together with the numerically computed DOS. There is visible disagreement for small $B_5$ regime seen in Fig. \ref{fig:B5}, as the traverse of the Fermi arc is the relatively large part of the trajectory. The linear dependence on the chemical potential is a simplistic approximation for the motion along Fermi arc, thus causing discrepancy.
The good agreement otherwise shows reliability of our model for predicting the influence of the external and pseudo-fields. Thus our prediction enable extraction of the values of $B_5$ as a function of strain applied to material by applying external magnetic field and measuring SdH or dHvA quantum oscillations.

We stress that results presented here apply both to time-reversal- and inversion-broken Weyl semimetals, since one can think of the latter as two time-reversed copies of the former. While locally the pseudo-CME might add up to a zero net contribution in time reversal symmetric systems due to the cancellation between time reversed pairs of nodes, the trajectories are still modified by them, and the effect on quantum oscillations should still be present. 

The case of Dirac semimetals is more subtle: it is known that the strain can develop spin-orbit coupling gapping out the Dirac semimetals like Cd$_3$As$_2$, as the symmetry protecting the cones is broken\cite{wang2012dirac,Wang2013}. Nevertheless, we predict that small $B_5$ is still accessible in the experiment in the limit of high magnetic field or high chemical potential with respect to Dirac point compared to spin-orbit gap. In the first case the two Weyl cones corresponding to the same Dirac cone have opposite spins, and are shifted in energy and momentum due to Zeeman term. Thus, we predict two sets of quantum oscillations corresponding to the two spin sectors to be present. In the second case the gap near the Dirac points does not influence the physics at high chemical potential and our predictions remain intact.

\acknowledgements The authors are indebted to inspiring discussions with Ady Stern, and would liked to thank Philip Moll, Adolfo Grushin, and Andrew Potter for useful comments. Numerical simulations were performed using Kwant code~\cite{groth2014kwant}.

\newpage

\begin{widetext}

\section{Supplementary material}

\subsection{Dispersion relation}

\begin{figure*}[t!]
\centering
\subfigure[]{
\includegraphics[width=5.5cm]{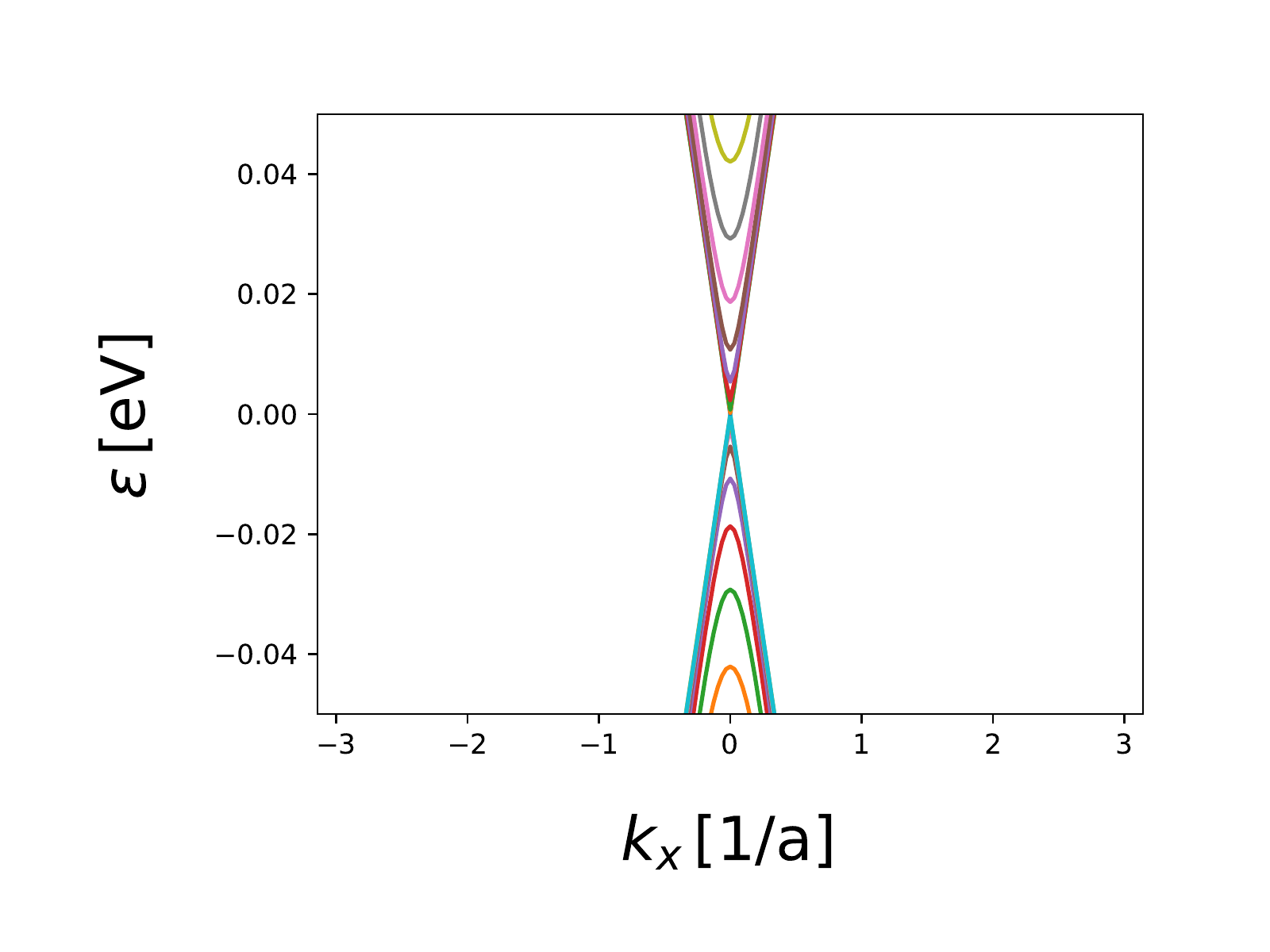}}
\subfigure[]{
\includegraphics[width=5.5cm]{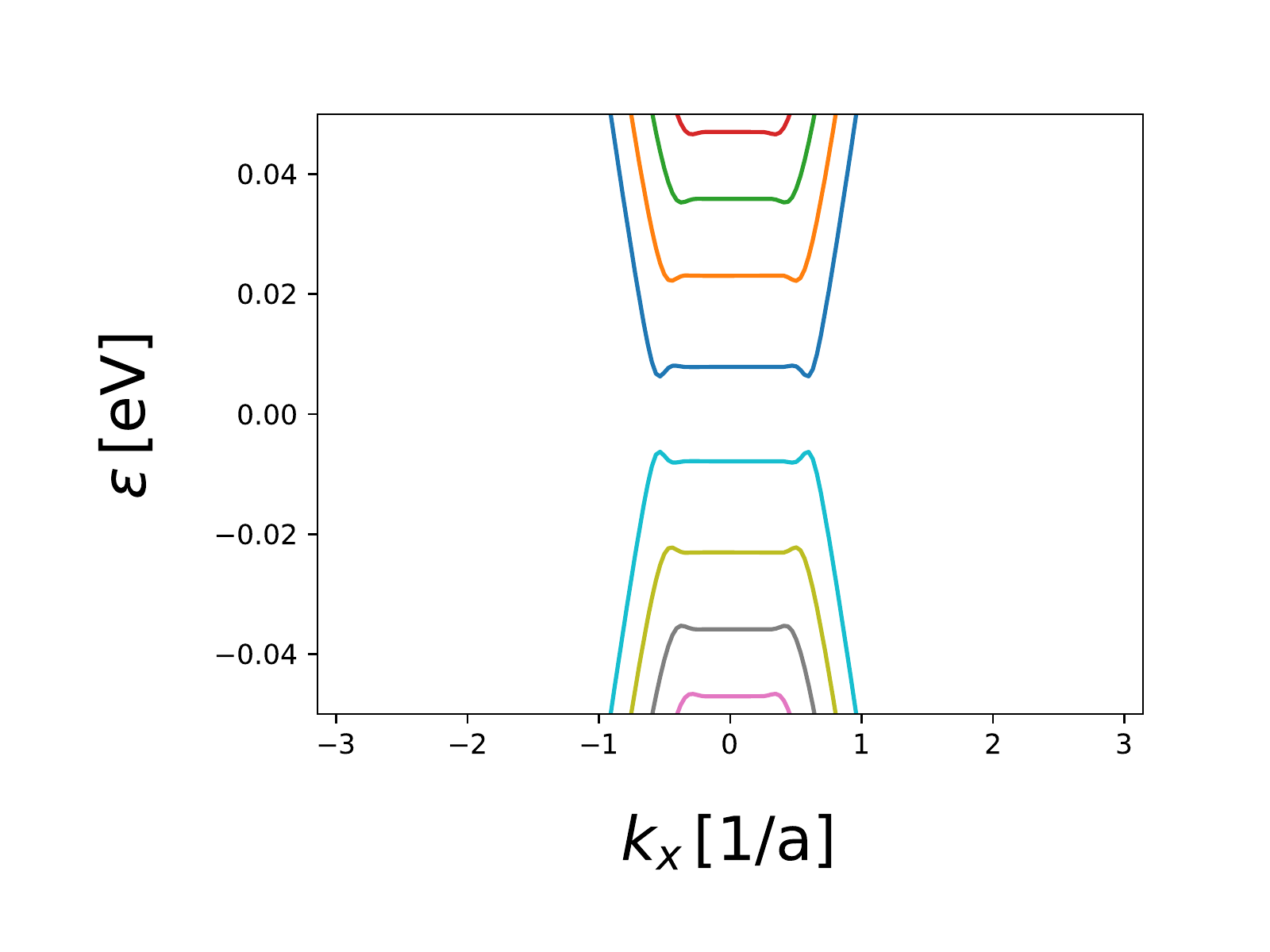}}
\subfigure[]{
\includegraphics[width=5.5cm]{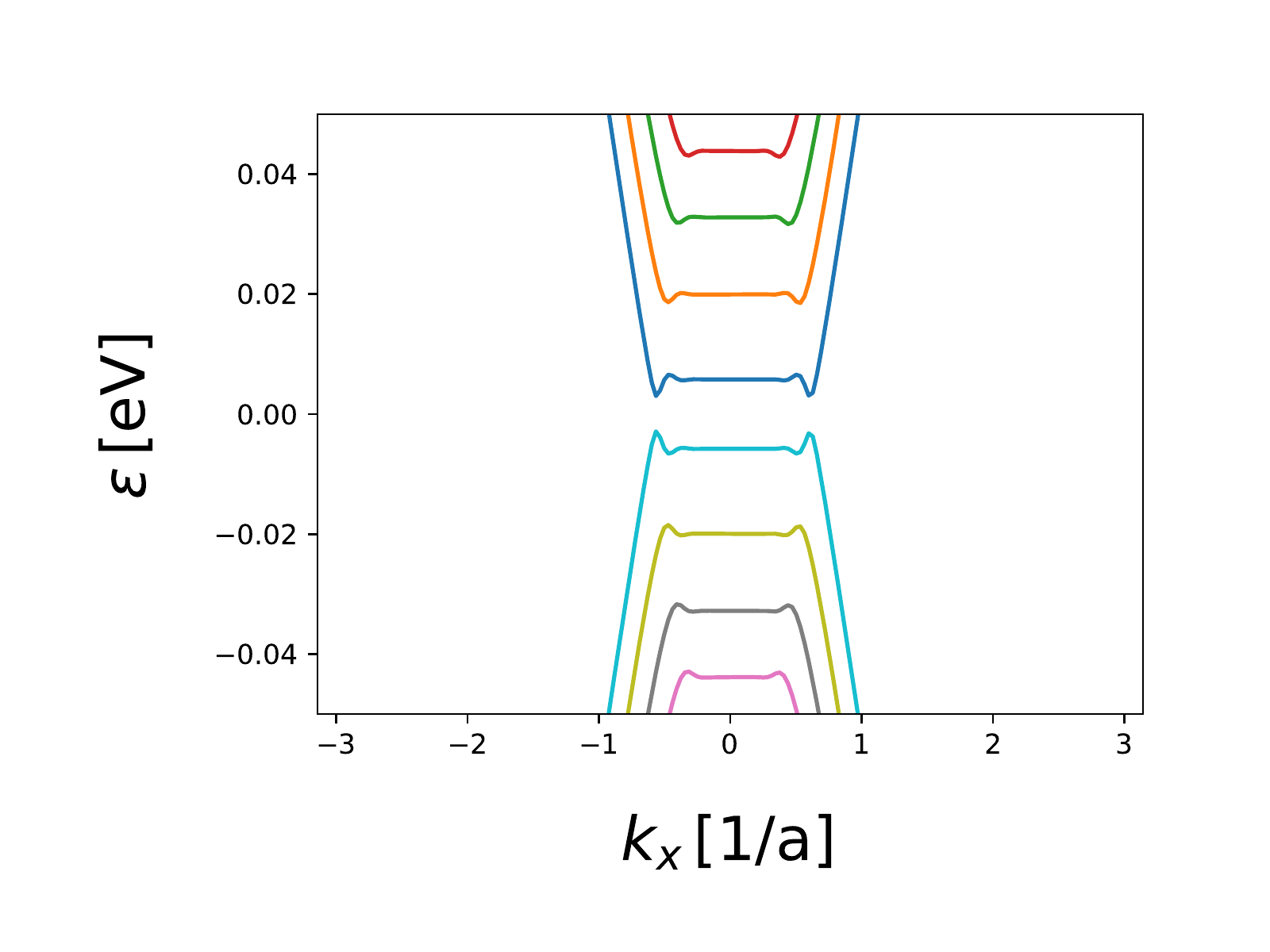}}
\caption{Dispersion relation along $k_x$ for $80\times 240$nm slab of the Weyl semimetal. (a): $B=B_5=0$T, (b): $B=0.5$T, $B_5=0$T, (c): $B=0.5$T, $B_5=0.25$T. The rest of the parameters are the same as in Fig. \ref{fig:B5}. (b) and (c) show Landau levels forming and the dependence of their energies on $B_5$.\label{fig:dispersion}}
\end{figure*}

To corroborate the findings of the main text, here we present the dispersion relations for $B=B_5=0$T; $B=0.5$T, $B_5=0$T; and $B=0.5$T, $B_5=0.25$T in Fig. \ref{fig:dispersion}(a), (b), and (c) correspondingly.
\end{widetext}
\subsection{Changes in the total arc length}
In this section we consider the generic case in which the total length of the Fermi arcs is modified by the existence of a bulk $B_5$. This can occur, for example, in strained samples where strain is applied on one surface and gradually relaxes to zero away from that surface such that the opposite side of the sample maintains the original unstrained value.  

Getting back to equation \eqref{semiclassic}, we note that both $L_{eff}$ and $S_k$ depend on $B$, $B_5$, and $L$. Writing out $S_k$ explicitly
\begin{eqnarray}
2\pi(n+\gamma)=&\mu(2L'/v+k_0(B_5)/evB) \nonumber \\ &+ k_0(B_5)\mu_0/evB.
\end{eqnarray}
where $k_0(B_5) = (k_0(0, B_5) + k_0(L, B_5)) / 2$, $k_0(z, B_5)$ is the Weyl node separation as a function of position in $z$ direction and $B_5$. This expression transforms to eq. \eqref{semiclassic} when $k_0(0) = k_0(L) = k_0$.

Let us now estimate the change in the overall arc length as follows. We assume that $B_5$ is uniform in the bulk of the sample, i.e the Weyl node separation changes linearly from one surface to another in the $z$ direction. Hence the total arc length is given by 
\begin{equation}
k_0(z, B_5)=k_0+B_5(z-z_0)
\end{equation}
where $z_0$ denotes the position in which the nodal separation is unperturbed. We set the sample position between $z=0$ and $z=L$, so that $k_0(0, B_5)=k_0-B_5z_0$ and $k_0(L, B_5)=k_0+B_5(L-z_0)$ and the change is the total length of the arcs can be estimated as $B_5(L-2z_0)$ and is linear in $B_5$ and $L$. 

Consequently, while we expect oscillations, they will not be at constant intervals as a function of $B$ or $B_5$. We can consider the two limits of $B_5/B\gg1$ and $B_5/B\ll1$. For the latter, $L\approx L'$ and we get that 
\begin{equation}
\Delta(1/B)=2\pi e v/k_0(B_5) (\mu_0+\mu)
\end{equation}
the expression is of a similar form to the one appearing in \cite{zhang2016quantum}, but now $k_0$ changes linearly in $B_5$. 
For $B_5\gg B$ we have that $L'\approx L B_5/B$ so that 
\begin{equation}
2\pi\hbar(n+\gamma)=\mu(2LB_5/v B+k_0(B_5)/evB)+k_0(B_5)\mu_0/evB
\end{equation}
and therefore 
\begin{eqnarray}
\Delta(1/B)=2\pi e v\left[2e\mu LB_5+k_0(B_5)(\mu+\mu_0) \right]^{-1}
\end{eqnarray}
And the denominator again changes linearly in both $B_5$, $L$.
\subsection{Unisotropic Fermi velocity}
In equation ~\eqref{WeylH} the Fermi velocity was taken to have a constant and isotropic value, $v$. In practice, the Fermi velocity might have a different value depending on the direction, hence we now extend the calculation to the case where $\vec{v}=(v_{\perp},v_{\perp},v_z)$. Since we choose the direction of the external field to be in the $z$ direction and perpendicular to the surface, we take the surface velocity to be $v_{\perp}$ as well. The modification introduced by the anisotropy affect equation \eqref{bulk}, where $v$ should be replaces by an effective velocity which is a combination of $v_\perp$ and $v_z$ weighted by the magnitude of $B$ and $B_5$, i.e
\begin{equation}
v_{b}=\sqrt{(\sin\theta v_\perp)^2+(\cos\theta v_z)^2}
\end{equation}
where $\theta=\tan^{-1}(B_5/B)\equiv \tan^{-1}(x)$.  $v_b$ can be written as 
\begin{equation}
v_{b}=[1+x^2]^{-\frac{1}{2}}\sqrt{x^2 v_\perp^2+v_z^2}
\end{equation}
hence equation ~\eqref{bulk} becomes
\begin{equation}
\int \textbf{p}\cdot \textbf{dr}=2L\mu\frac{1+x^2}{\sqrt{x^2 v_\perp^2+v_z^2}}
\end{equation}
The quantization condition \eqref{semiclassic} then becomes
\begin{equation}
2\pi(n+\gamma)=2L\mu\frac{1+x^2}{\sqrt{x^2 v_\perp^2+v_z^2}}+\frac{k_0(B_5)(\mu+\mu_0)}{v_\perp eB}
\end{equation}
As expected, when $v_\perp$ and $v_z$ are comparable, the same analysis that is presented in the main text holds. Alternatively, if one of these velocities is much larger than the other, the result depends on the magnitude of $x$ and can show different dependencies on $x$. For example, if $v_\perp\gg v_z$ and $x\gg 1$, we find that 
$$\frac{1+x^2}{\sqrt{x^2 v_\perp^2+v_z^2}}=\frac{1+x^2}{v_\perp\sqrt{x^2+(v_z/v_\perp)^2}}\approx \frac{x}{v_\perp}$$
while if $x\ll 1$ we have 
$$\frac{1+x^2}{\sqrt{x^2 v_\perp^2+v_z^2}}=\frac{1+x^2}{v_\perp\sqrt{x^2+(v_z/v_\perp)^2}}$$
The result in this case clearly depends on the hierarchy of $x$ and $v_z/v_\perp$. For $x\gg v_z/v_\perp$ one gets $1/(v_\perp x)$, while for $x\ll v_z/v_\perp$ the result is $(1+x^2)v_\perp/v_z$ as the lowest order correction in $x$.

\end{document}